\documentclass[10pt, conference]{IEEEtran}
\usepackage{cite}
\usepackage{blindtext}
\usepackage{amsmath,amssymb,amsfonts}
\usepackage[ruled,linesnumbered]{algorithm2e}
\usepackage{graphicx}
\usepackage{textcomp}
\usepackage{algpseudocode}
\usepackage{xcolor}
\usepackage{subfig}
\usepackage{booktabs}
\usepackage{xurl}
\usepackage{hyperref}
\usepackage{array}
\usepackage{diagbox}

\makeatletter
\newcommand{\linebreakand}{%
  \end{@IEEEauthorhalign}
  \hfill\mbox{}\par
  \mbox{}\hfill\begin{@IEEEauthorhalign}
}
\makeatother

\IEEEoverridecommandlockouts
\def\BibTeX{{\rm B\kern-.05em{\sc i\kern-.025em b}\kern-.08em
    T\kern-.1667em\lower.7ex\hbox{E}\kern-.125emX}}
\begin{document}
\title{Detecting Stimuli with Novel Temporal Patterns to Accelerate Functional Coverage Closure}

\author{\IEEEauthorblockN{Xuan Zheng}
\IEEEauthorblockA{\textit{Trustworthy Systems Laboratory} \\
\textit{University of Bristol}\\
Bristol, UK \\
dq18619@bristol.ac.uk}
\and
\IEEEauthorblockN{Tim Blackmore}
\IEEEauthorblockA{\textit{Verification Team} \\
\textit{Infineon Technologies}\\
Bristol, UK \\
Tim.Blackmore@infineon.com}
\and
\IEEEauthorblockN{James Buckingham}
\IEEEauthorblockA{\textit{Verification Team} \\
\textit{Infineon Technologies}\\
Bristol, UK \\
James.Buckingham@infineon.com}
\linebreakand

\IEEEauthorblockN{Kerstin Eder}
\IEEEauthorblockA{\textit{Trustworthy Systems Laboratory} \\
\textit{University of Bristol}\\
Bristol, UK \\
kerstin.eder@bristol.ac.uk}
}

\maketitle    

\begin{abstract}
Novel test selectors have demonstrated their effectiveness in accelerating the closure of functional coverage for various industrial digital designs in simulation-based verification. The primary advantages of these test selectors include performance that is not impacted by coverage holes, straightforward implementation, and relatively low computational expense. However, the detection of stimuli with novel temporal patterns remains largely unexplored. This paper introduces two novel test selectors designed to identify such stimuli. The experiments reveal that both test selectors can accelerate the functional coverage for a commercial bus bridge, compared to random test selection. Specifically, one selector achieves a 26.9\% reduction in the number of simulated tests required to reach 98.5\% coverage, outperforming the savings achieved by two previously published test selectors by factors of 13 and 2.68, respectively.

\end{abstract}


\section{Introduction}\label{s:introduction}
Verification is a critical step in the development process of digital designs, ensuring that the functionality of a design complies with its specifications. However, as digital designs grow in functional complexity, verification has become more time-consuming and requires greater engineering efforts. For example, simulation-based verification, a major technique in functional verification, has become less efficient for large-scale designs. One factor contributing to this inefficiency is the simulation of randomly selected tests, which yields diminishing functional coverage gains as verification progresses. Consequently, redundant tests continuously consume verification resources, such as computing power, while verifying very few new functionalities. This redundancy slows down verification, compelling engineers to generate more tests in an attempt to verify the remaining complex functionalities. Still, random test selection remains the state-of-the-art technique used in the industry.

In recent years, prioritizing the simulation of novel tests has been shown to significantly accelerate coverage closure for designs of commercial complexity, irrespective of the number of coverage holes, compared to random test selection~\cite{bworld, liang2023late, zheng2023using}. It is posited that simulating the un-simulated tests that are novel compared to those already simulated is more likely to hit new coverage areas~\cite{guzey2008functional}. Unlike most Machine Learning (ML) algorithms used in Coverage-Driven Verification (CDV) that empirically learn the correlation between stimuli or the constraints of stimuli generation, and coverage products~\cite{araiza2015coverage, yusurvey}, capturing novel tests does not require positive examples for infrequent coverage products. Thus, these test selectors are scalable to industrial-level designs. However, previous test selectors have not discussed capturing the novel temporal dependencies between sequentially-driven stimuli. The order in which stimuli are delivered to a Design under Verification (DUV) is highly correlated with the intricate and corner-case functionalities of various digital designs, such as CPUs and bus bridges.

This paper explores the feasibility of using two deep learning (DL) models to identify stimuli with novel temporal patterns. The proposed DL models are the encoder of a transformer~\cite{vaswani2017attention} and a long short-term memory (LSTM) autoencoder~\cite{8735594}, both of which are designed for processing time-series data. Nevertheless, the implementation of them in this paper enables them to detect not only the temporal patterns but also novel attribute values of a stimulus, or both types of novelty. The effectiveness of these DL models is evaluated by the simulation savings achieved to reach several coverage goals in a functional coverage model defined for a commercial bus bridge.

From the experiments conducted, it is evident that both DL models substantially accelerate coverage progress compared to random test selection. Specifically, the transformer encoder achieves a reduction in simulation efforts by 4.46\% to attain a 98.5\% coverage goal, while the LSTM autoencoder further reduces these efforts by 26.9\% for the same goal. In addition, the actual simulation time saved by these two test selectors amounts to 41.61 and 254.53 hours, respectively, on a single EDA license, even after accounting for the additional computational expenses they introduce. Notably, the simulation saving realized by the LSTM autoencoder to reach a 98.5\% coverage level is 13 and 2.68 times greater than those achieved by the previous novel test selectors~\cite{bworld,zheng2023using} and~\cite{liang2023late} that have demonstrated distinguished performance across several industrial designs.

Furthermore, our test selectors still inherit the advantages of previous test selectors:
\newline

1. Highly automated: While the domain knowledge of a DUV can be integrated into our test selector, it is noteworthy that the embedded knowledge is relatively basic and limited in scope. As a result, the construction of our test selector remains straightforward and uncomplicated. The generated tests can be automatically encoded into input vectors driven to the test selectors.

2. Easy to be integrated into a verification environment: Generated tests are only ordered for simulation.

3. Their performance is not influenced by the number of infrequent coverage products.

4. Relatively light computational expense: Compared to the simulation savings facilitated by the test selectors, the extra expenses brought by them are relatively low .
\newline

The paper is structured as follows: Section~\ref{s:previous} reviews and compares related works to identify the research gap inspiring this work. Section~\ref{s:Metho} introduces our methodology. Section~\ref{s:Experimental} presents and discusses the experimental setup, results, and the comparison between each method. Finally, Section~\ref{s:Conclusion} concludes the paper and discusses future work.

\section{Previous Work}\label{s:previous}

In CDV, ML algorithms have been extensively researched for empirically learning the correlation between the constraints used to bias the generation of tests or tests themselves driven to a DUV and their associated coverage products~\cite{araiza2015coverage,yusurvey}. However, for missing and rare coverage products, the associated positive tests remain scarce. Consequently, ML models struggle to effectively establish correlations between infrequent coverage products and the parameters extracted from test generators or tests.

On the other hand, a novel test selector was proposed in~\cite{guzey2008functional} to prune redundant tests during the simulation, regardless of the number of infrequent coverage products. It is assumed that tests novel to those already simulated more likely lead to new coverage. Under this assumption, identifying novel tests and prioritizing their simulation could contribute to efficient coverage closure. The effectiveness of novel test selectors has been proven in the simulation-based verification environments of several commercial designs~\cite{bworld, liang2023late, zheng2023using}.

In~\cite{liang2023late}, a range of ML models, including an autoencoder, an isolation forest, and a one-class support vector machine, are employed to develop novel test selectors. The inputs to these test selectors are not derived from generated tests or the attributes of the constraints to bias test generation. Instead, a test is simulated on a functional equivalence model first, and then the associated structural coverage obtained in the equivalence model is extracted as the input. The authors found that infrequently-hit RTL coverage products are highly correlated with the rare products in the structural coverage model. Additionally, test simulation on a functional equivalence model is 10 to 100 times faster than that on RTL. The DUVs used in the experiment are two datapath-oriented units from an industry GPU. Around 20,000 tests are generated for each unit. The number of functional coverage products for unit A and unit B are 17,520 and 3,099, respectively. To reach a 90\% coverage level, 75\% of the simulation time can be saved for Unit A, while that saving in Unit B is 85\%. However, not all verification environments have a functional equivalence model for a DUV. Even if a functional equivalence model exists, collecting structural coverage can still be restricted by the programming language used to write the model. Therefore, representing generated stimuli as inputs to novel test selectors is a more comprehensive approach that suits the verification environments of various DUVs.

The test selector in~\cite{bworld} is based on an autoencoder with generated tests directly encoded as the input to the selector. The test selector is trained to reproduce input tests, and those with high reconstruction error are deemed novel. A commercial signal processing unit (SPU) in an ADAS system was chosen to verify the effectiveness of the test selector. The functional coverage model contains about 6,000 white-box products. More than 85,000 tests are generated, and each test consists of values from functionality configuration registers. From the results, 60\% of simulation time can be saved to reach the 99.5\% coverage level, compared to random selection.

In~\cite{zheng2023using}, a neural network-based test selector with a configurable output space is proposed. Three configurations are mentioned: autoencoder, density, and coverage-novelty configurations, with the latter two being supervised algorithms. For the density configuration, each output neuron is configured to output the probability of a coverage product being hit by an input test. Instead of relying on the prediction of whether a coverage product will be hit by an input test, the algorithm uses hidden neurons as novelty estimators to estimate how novel an input test is relative to the simulated tests (training data). For the coverage-novelty configuration, the algorithm estimates how novel an input test is in the coverage space formed by the covered products. Thus, the lack of positive examples is subtly circumvented by both supervised configurations. The experimental data used is the same as in~\cite{bworld}, except the number of functional coverage products has been increased to over 8,000. The results reveal that all three configurations outperform random test selection, with the autoencoder demonstrating dominant performance.

Previous test selectors have overlooked the identification of stimuli with novel temporal patterns, such as novel dependencies between transactions sequentially driven to a DUV. In~\cite{gogri2020machine,9806210}, pruning redundant sequential stimuli by ML models for efficient simulation has been researched. However, these models are implemented to predict whether coverage products can be activated by generated tests. As a result, the lack of positive examples for infrequent coverage products still remains a significant obstacle to prevent this method being implemented in the verification environment of large-scale designs.

\section{Methodology}\label{s:Metho}

\subsection{Operation of Novel Test Selectors}
\subsubsection{General Operation Scheme}

The operation of novel test selectors in the environment of simulation-based verification has been elucidated in~\cite{zheng2023using}. A small number of initial tests are chosen to warm up a test selector. Subsequently, the iteration of novel test selection begins, wherein all previously simulated tests are used to train a test selector, and then a batch of un-simulated tests, deemed most novel compared to the simulated ones, are selected for the next round of simulation. This iteration process terminates when either a termination criterion is reached, or all generated tests have been simulated.

\subsubsection{Sampling Inputs for Novel Test Selectors}
Depending on the domain knowledge of a DUV, a generated test can be represented in its entirety as an input to test selectors (a coarse-grained input), or it can be decomposed into smaller components. Each component is then treated as an input to test selectors (a fine-grained input).

In~\cite{bworld,zheng2023using}, the DUV is a highly-configurable SPU. A generated test encompasses the values of all configuration registers that configure the functionality of the SPU. The operation of the SPU under a given functional mode requires a specific combination of values set for all the registers in advance. Consequently, an entire test is sampled as an input driven to the test selectors, with each attribute of the input corresponding to a configuration register. In this scenario, the mission of test selectors is to identify tests with novel attribute values.

\begin{figure}[!t]
  \centering
  \subfloat[Coarse-grained Inputs]{\includegraphics[width=1\linewidth]{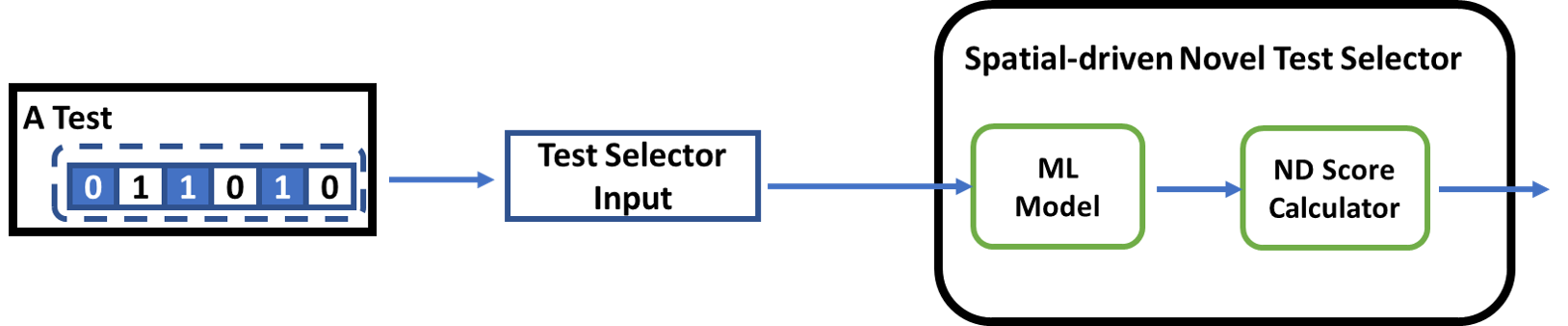}\label{fig:f1}}
  \hfill
  \vspace{1cm}
  \subfloat[Fine-grained Inputs]{\includegraphics[width=1\linewidth]{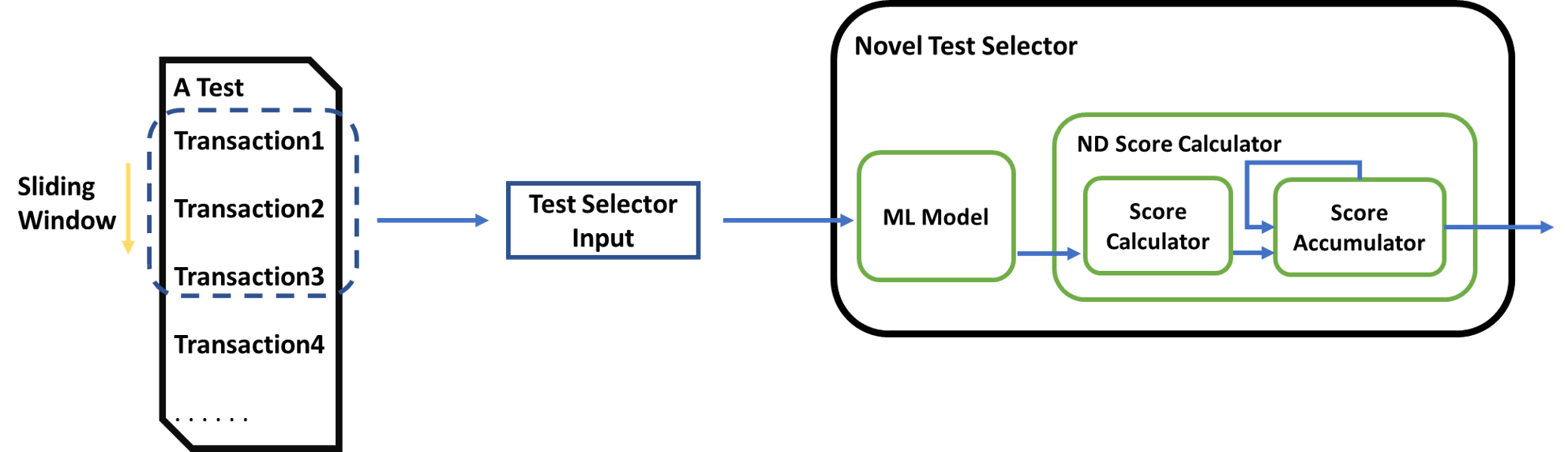}\label{fig:f2}}
  \caption{Granularity Difference of the Inputs to Test Selectors}
  \label{fig:selector_diff}
\end{figure}

For designs characterized by sequentially-driven inputs, such as bus bridges, a generated test typically contains more transactions than the maximum number the design can process in a pipeline. Furthermore, the functionalities of these designs are not only related to the combinations of attribute values of a transaction but also to the temporal dependencies between transactions being processed. For example, a functional coverage product for a bus bridge can relate to the types of transactions (Write/Read) and the source and destination dependencies among a sequence of transactions. Thus, to effectively capture both value novelty and sequential-dependency novelty, the input to test selectors should be a sequence of transactions rather than an entire test. This granularity difference in constructing the inputs for test selectors is depicted in Figure~\ref{fig:selector_diff}.

The granularity degree of inputs of previous test selectors is depicted in Figure~\ref{fig:f1}, while that in our test selectors is illustrated in Figure~\ref{fig:f2}. In our test selectors, a sliding window is used to sample a sequence of fine-grained inputs. To determine the size of the sliding window, the domain knowledge of a DUV can be referred to. For example, the window size can be set to the maximum depth of the pipeline in a bus bridge. The number of steps to slide the window is also a user-defined parameter. An efficient scheme is to set the number of sliding steps equal to the window size, ensuring that each input sequence does not overlap with another. In such a way, expensive training time can be avoided with fewer input sequences. All transactions within a test are sampled. Therefore, a generated test may comprise multiple inputs that are driven to test selectors. 
\subsubsection{Construction of Novelty Scores}
As the atomic unit of simulation is a test, the novelty scores of sampled sequences need to be effectively gathered to form the final novelty score for a test. Tests with higher novelty degrees should be prioritized during simulation because they are more likely to activate new functionalities.

The method to gather the novelty scores of sampled sequences is described in Equation~\ref{eq:ND_Score_Gathering}.

\begin{subequations}
\vspace{0.2cm}
\label{eq:ND_Score_Gathering}
\begin{equation}
 S_{seq} = {\sum_{p=1}^{L}RE_p/L}
 \label{eq:ND_Score_Seq}
\end{equation}    
\vspace{0.2cm}
\begin{equation}
 S_{test} = \sum_{j=1}^{n}(S_{{seq}_j})^2/n
 \label{eq:ND_Score_Test}
\end{equation}
\end{subequations}
\vspace{0.2cm}

For a sequence with L number of transactions, the novelty score is calculated as $S_{seq}$. Here, $RE_p$ is the novelty score of the $p$th transaction in a sequence, which is calculated by novel test selectors. $S_{test}$ is the novelty score assigned to a test, where n sequences are sampled from the test.

Simply averaging the novelty scores of all sampled sequences could lead to a problematic scenario where a test with a large portion of slightly novel sequences is assigned a higher novelty score than a test with a small portion of significantly more novel sequences. To address this issue, the novelty score of each sequence is squared in Equation~\ref{eq:ND_Score_Test} where $n$ is the total number of sampled sequences in a test.

\subsection{DL Models of Novel Test Selectors}\label{ML_Models}

The DL models utilized in our test selectors are the encoder of a transformer and an LSTM autoencoder. Both DL models are trained to reconstruct a sequence of input transactions, with mean squared error (MSE) used to calculate the reconstruction error $RE$, as shown in Equation~ref{eq:MSE}. Here, $n$ represents the number of attributes of an input transaction. $I_j$ is the value of the original attribute $j$, and $O_j$ denotes the value of the reconstructed attribute $O_j$. 

\vspace{0.2cm}
\begin{equation}
  \label{eq:MSE}
  \begin{aligned}
   RE &= \frac{1}{n}\sum_{j=1}^{n}(I_j - O_j)^2
  \end{aligned}
\end{equation}
\vspace{0.2cm}

\begin{figure}
    \centering
    \includegraphics[width=1\linewidth]{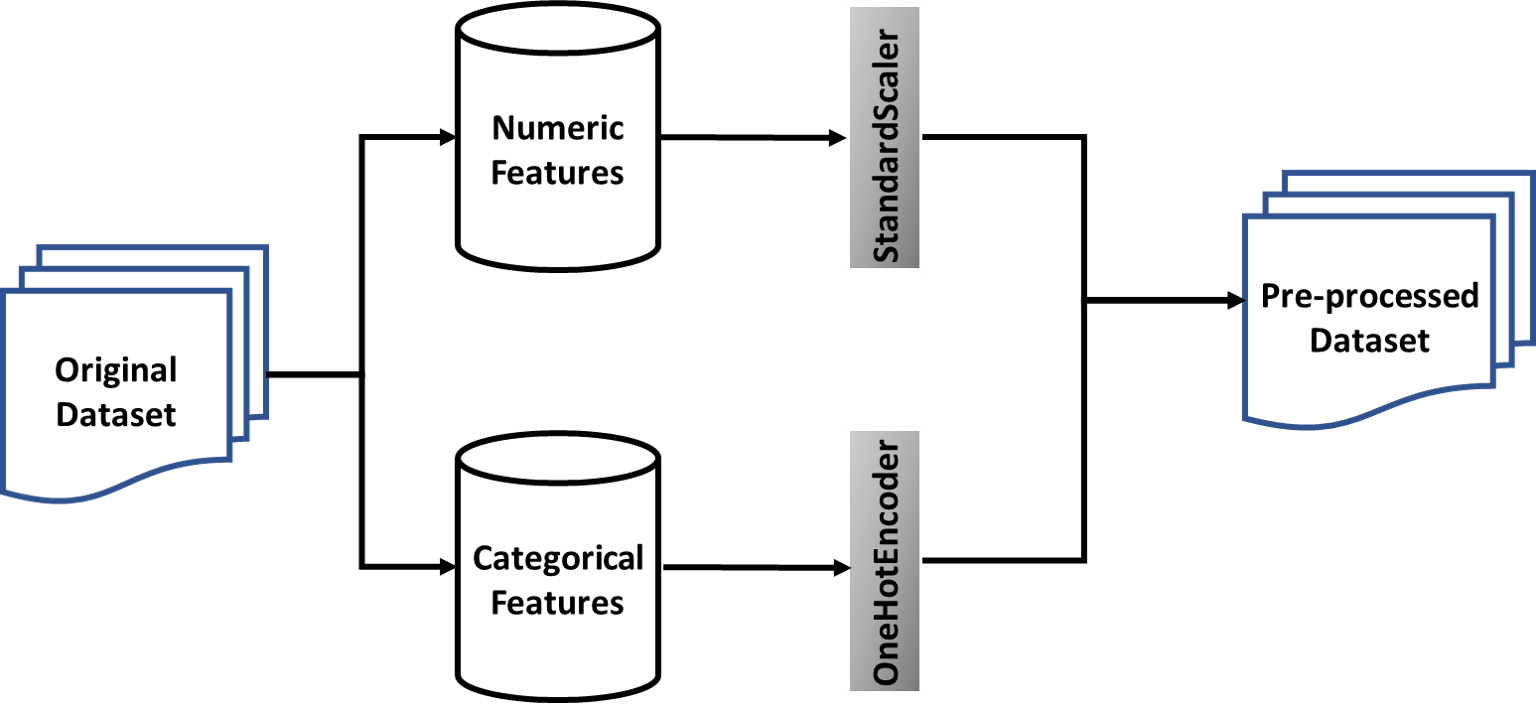}
    \caption{Pre-processing Sampled Inputs}
    \label{fig:f3}
    \vspace{1.0cm}
\end{figure}

A sequence of sampled transactions needs to be pre-processed to make them suitable for the operation of DL models. This process is illustrated in Figure~\ref{fig:f3}.

For the numeric attributes of transactions, they should be standardized to ensure that no single attribute disproportionately dominates the representation of input data and thus biases the learning of DL models. Standardization involves removing the mean from an attribute and then scaling it to unit variance. Categorical attributes are one-hot encoded to ensure that each category of an attribute has the same importance as another.

\begin{figure}
    \centering
    \includegraphics[width=1\linewidth]{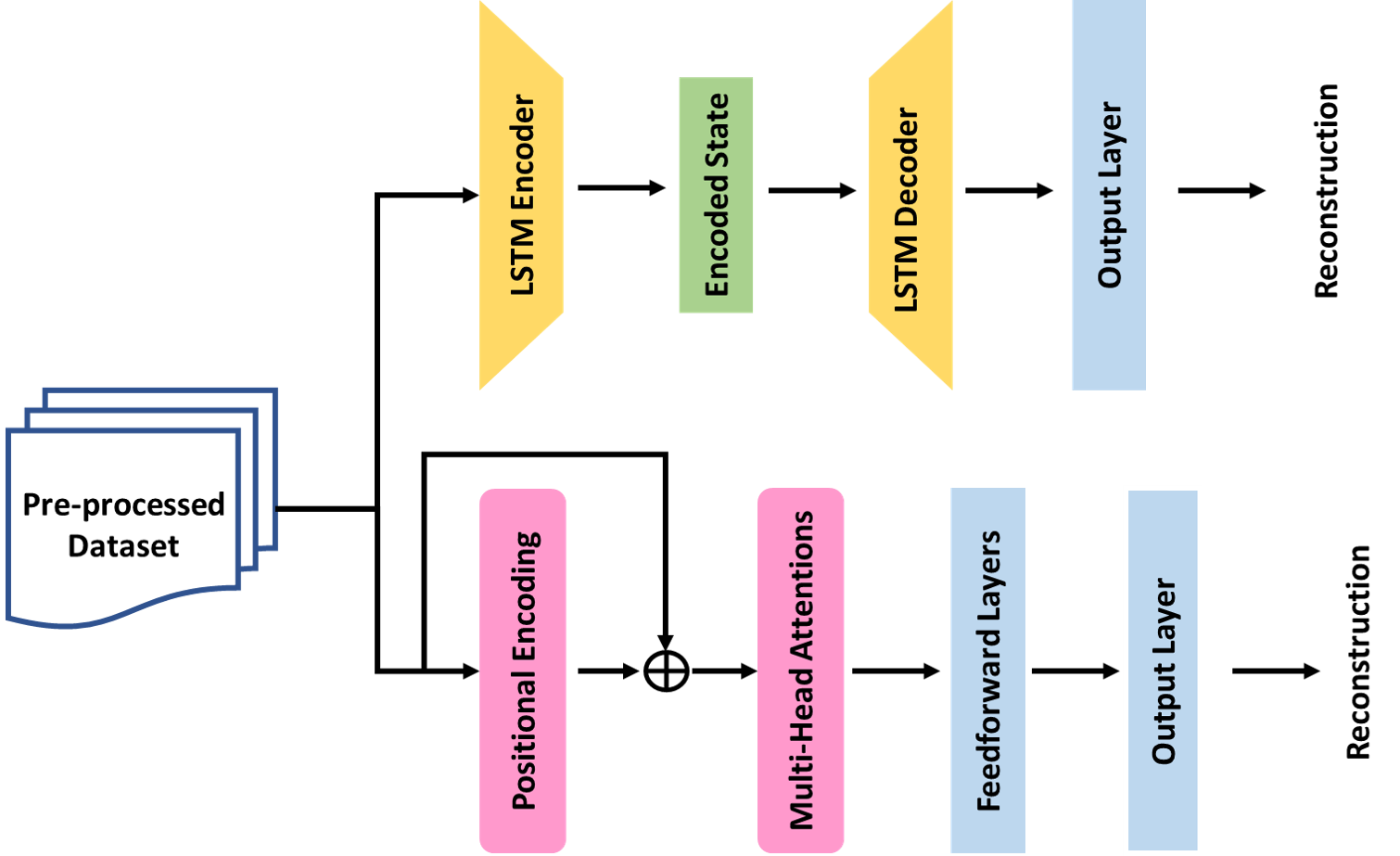}
    \caption{DL Models}
    \label{fig:f4}
\end{figure}

Afterwards, the pre-processed data is fed into the two DL models depicted in Figure~\ref{fig:f4}. The model in the top half of the figure is the LSTM autoencoder. LSTM, a type of Recurrent Neural Network (RNN), incorporates feedback connections, enabling it to process entire sequences of data rather than individual data points. The backpropagation training of RNNs through time can lead to an exploding or vanishing gradient problem, which LSTM units mitigate by employing several gates~\cite{shewalkar2019performance}.

The model in the bottom half of the figure is the encoder of a transformer. Key factors enabling the Transformer to handle time-series data include positional encoding and the attention mechanism. Positional encoding provides the model with information about the relative or absolute position of transactions in the sequence. In this paper, we adopt sine and cosine functions, as utilized in~\cite{vaswani2017attention}, to embed positional information. Attention mechanisms allow the transformer encoder to model the dependencies between transactions regardless of their distance in the sequences of transactions. The dot-product attention mechanism is employed in our DL model.

\section{Experimental Evaluation}\label{s:Experimental}
\subsection{Experiment Background}\label{s:Exp_background}
In this case study, the on-chip bus bridge based on the Shared Resource Interconnection (SRI) of the TriCore system is chosen as the DUV. SRI serves as the high-speed system bus and has high bandwidth~\cite{OMS3}. The central module of the interconnect is SRI crossbar, which connects all components within one SRI system. The SRI crossbar module handles, arbitrates, and forwards the communication between all connected SRI\_Master and SRI\_Slave peripherals. Additionally, SRI crossbar supports parallel transactions between different SRI\_Master and SRI\_Slave peripherals, as well as pipelined requests from SRI\_Master interfaces and pipeline address phases to connected SRI\_Slave interfaces.

The dataset used for the experiments comprises two parts: generated tests and their corresponding coverage information. This dataset is obtained from the simulation-based verification environment of the SRI bus bridge. It includes 2,768 golden tests in addition to 10,000 random tests. Each test contains 795 bus transactions on average, with the minimum number of transactions being 471 and the maximum number being 1092. From the pipeline, parallelism, and data-pacing coverage models defined for the bus bridge, 12,320 cross-coverage products are extracted. The data-pacing coverage model illustrates the combinations of cycles that each data beat of a burst transaction waits to be processed. These three coverage models depict the intricate temporal characteristics of the bus design.

A bus transaction originally includes 5 categorical attributes and 10 numeric attributes. The numeric attributes are standardized using the library in Scikit-Learn~\cite{scikit-learn}, while the categorical attributes are encoded using the OneHotEncoder in Scikit-Learn~\cite{scikit-learn}. In total, 104 pre-processed attributes are obtained.

\subsection{Experiment Setup}\label{s:Exp_setup}

The following test selection methods are compared:
\newline

RD: Random test selection, the state-of-the-art technique commonly deployed in simulation-based verification for commercial designs.

AE: Traditional autoencoder-based novel test selector, as utilized in existing literature~\cite{bworld,zheng2023using,liang2023late}.

IF: Isolation forest-based novel test selector, as utilized in existing literature~\cite{liang2023late}.

TE: Transformer encoder-based novel test selector, proposed in this paper.

LSTM: LSTM autoencoder-based novel test selector, also proposed in this paper.
\newline

The construction of smart test selectors utilizes the PyTorch ~\cite{NEURIPS2019_9015} and Scikit-Learn~\cite{scikit-learn} libraries. Each neural network-based selector in the experiment is configured with a proportional structure, wherein the number of neurons in each hidden layer is either twice, half, or the same as the number in the previous layer.

The window size to sample input sequences is set to 3, resulting in 3 transactions being driven to a test selector each time. This decision is based on the maximum depth of the pipeline in the SRI bus bridge, which is 3. Additionally, the number of steps to slide the window is also set to 3. For example, if a test contains 6 transactions {$T_1$, $T_2$, $T_3$, $T_4$, $T_5$, $T_6$}, the first sampled sequence comprises the group of transactions {$T_1$, $T_2$, $T_3$}, and the second sequence contains transactions {$T_4$, $T_5$, $T_6$} after the window is slid by 3 transactions. The rationale for having non-overlapping sequences is to reduce the volume of data processed by the test selectors, thereby minimizing the computational expense incurred by test selectors.

The number of initial tests used to warm up each test selector is set to 50. Subsequently, 4000 tests are selected for simulation in each selection iteration. Retraining occurs after the simulation of the selected tests, with all simulated tests being used to form the retraining set. To mitigate the risk of obtaining coincidental results, the experiment for each test selection method is repeated 10 times with different randomly sampled initial tests. The average result of 10 runs is presented for each selection technique. The training and inference of smart test selectors are conducted on a NVIDIA A30 GPU.

\subsection{Experimental Results}\label{s:result}

\begin{figure}[!t]
  \centering
  \subfloat[50 to 3050 simulated tests]{\includegraphics[scale = 0.35]{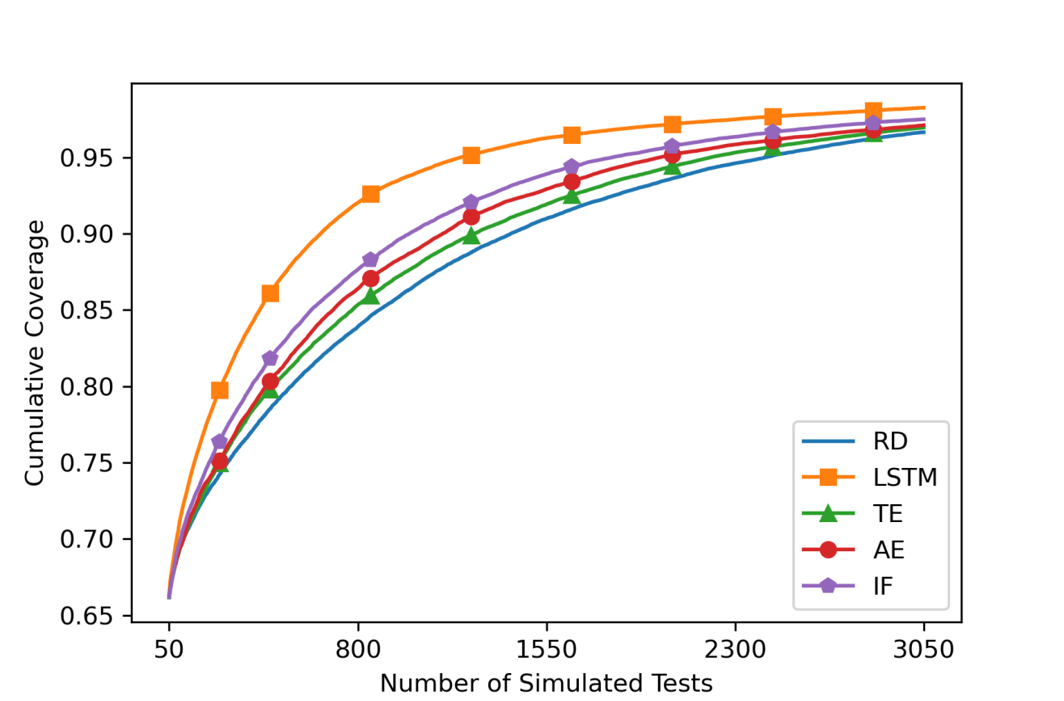}\label{fig:f5}}
  \hfill
  \vspace{0.1cm}
  \subfloat[3000 to 6000 simulated tests]{\includegraphics[scale = 0.35]{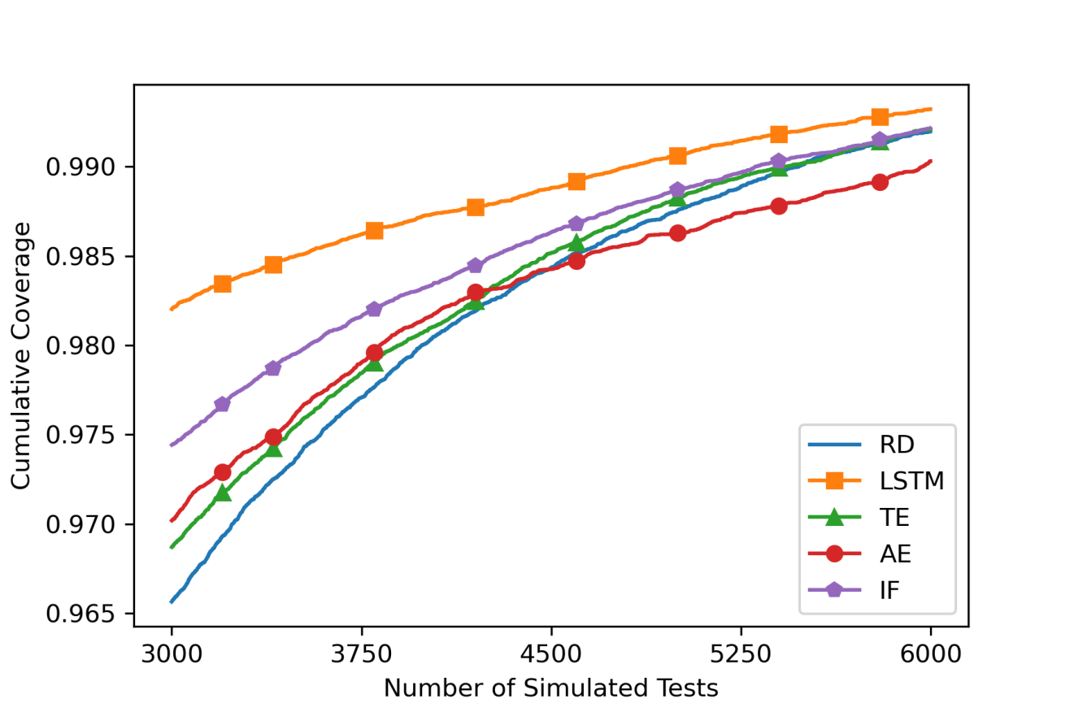}\label{fig:f6}}
  \caption{Coverage Progress v.s. Number of Simulated Tests}
  \label{fig:CovPlot}
\end{figure}

The coverage progress throughout the simulation is illustrated in Figure~\ref{fig:CovPlot}. As shown in Figure~\ref{fig:f5}, all the four smart test selectors can accelerate the coverage progress compared to the RD method during the initial stage of simulation. As the coverage levels exceed 96.5\%, the simulation efforts saved by TE and AE begin to decrease. The savings brought by IF also gradually diminish towards the end of the simulation. However, LSTM can still maintain notable simulation savings. The performance of AE is even worse than RD after reaching around the 98.3\% coverage level. This observation suggests that AE is less robust in capturing novel temporal relations between transactions than the other three smart test selectors.

\begin{table*}[tbhb]
\centering
\caption{Number of Simulated Tests to Achieve Given Coverage Goals}
\label{tab:Savings_Coverage_Goals}
\begin{tabular}{@{}lrrrrrrrrr@{}}
\toprule
\backslashbox{Goals}{Methods}
 
       & RD & AE & Savings & IF & Savings  & TE & Savings & LSTM & Savings \\ \midrule
95\%   & 2416   & 1992     & 424 (17.55\%)     & 1812       & 604 (25.00\%)        & 2166     & 232 (9.60\%)       & 1209             & 1207 (49.96\%)        \\
97\%   & 3301   & 2971     & 330 (10.00\%)    &  2644        & 657(19.90\%)        & 3036     & 220 (6.66\%)       & 1916             & 1385 (41.96\%)       \\
98\%   & 4087   & 3816     & 271 (6.63\%)    &   3533      &   554(13.56\%)     & 3882      & 159 (3.89\%)       & 2769             & 1318 (32.25\%)      \\
98.5\% & 4735   & 4637     & 98 (2.07\%)      &  4260       &  475 (10.03\%)    & 4513       & 211 (4.46\%)       & 3461             & 1274 (26.90\%)        \\
99\%   & 5664   & 5942     & -278 (-4.90\%)    &    5303      & 361 (6.37\%)      & 5483    & 144 (2.54\%)       & 4852             & 812 (14.34\%)       \\ \bottomrule
\vspace{0.3cm}
\end{tabular}
\end{table*}

Table~\ref{tab:Savings_Coverage_Goals} provides a quantitative comparison for five high coverage goals. The "Savings" column denotes the simulation savings of a smart selector relative to RD, while the other columns represent the numbers of simulated tests required to reach the defined coverage goals. The LSTM method exhibits the largest savings across all the five coverage goals. Notably, at the 98.5\% coverage level where directed testing is deployed to address remaining coverage holes in the practical project, LSTM achieves 13 times the savings of AE and 2.68 times the savings of IF. 

\begin{table}[tbhb]
\caption{Computational Expenses and Net Savings in Hours} 
\centering 
\begin{tabular}{|l|c|c|c|c|c|} 
\hline\hline 
Extra Expenses & 95\%  & 97\%  & 98\% & 98.5\% & 99\% \\ [0.05ex] 
\hline 
AE & 0.26 & 0.26 & 0.26 & 0.48 & 0.48\\ 
RF &  0.27 & 0.27 & 0.27 & 0.43 & 0.43\\
TE &  0.27 & 0.27 & 0.27 & 0.59 & 0.59\\
LSTM & 0.27 & 0.27 & 0.27 & 0.27 & 0.56\\ [1ex] 
\hline 
\noindent
Net Savings & 95\%  & 97\%  & 98\% & 98.5\% & 99\% \\ [0.05ex] 
\hline 
AE & 85.54 & 49.07 & 53.94 & 19.12 & -56.08 \\ 
RF &  120.53 & 131.13  & 110.53 & 94.57 & 71.77\\
TE & 46.13 & 43.73 & 31.53 & 41.61  & 28.21\\
LSTM & 241.13 & 276.73 & 263.33 & 254.53 & 161.84 \\ [1ex] 
\hline 
\end{tabular}
\label{table:NetSavings} 
\end{table}

For designs of commercial complexity, simulating a test can incur significant costs. Therefore, even minor percentages of saved simulated tests contribute significantly to reducing simulation time. This is highlighted in Table~\ref{table:NetSavings}, where the average time to simulate a test on the SRI bus bridge is 12 minutes on a single EDA license and an Intel Xeon Gold 6248R CPU. Net Savings are obtained by subtracting the computational expense from the saved simulation time.

\section{Conclusion}\label{s:Conclusion}

Novel test selectors have proven effective in reducing simulation redundancy for various commercial digital designs. However, the identification of novel temporal dependencies between sequentially-driven stimuli remains an unexplored area. This paper proposes two test selectors designed to identify tests with both novel temporal patterns and attribute values. Similar to previous test selectors, ours are scalable to industrial-level designs due to their ease of implementation, relatively light computation, and performance not being impacted by coverage holes. In our experiments, we compare the performance of our two test selectors with another two from the literature using a commercial bus bridge as the DUV. Our models can substantially accelerate coverage progress compared to random selection throughout the test simulation. Moreover, one of our test selectors significantly outperforms the previous methods. This suggests that our test selector is more adept at capturing novel sequential stimuli.

Future work will explore combining novel test selectors with Reinforcement Learning (RL) to automatically bridge the gap between test generation and coverage closure. The novelty score could be a proxy for coverage, enabling RL agents to explore different actions and states in a faster neural network-based environment and thus reducing the need for expensive test simulation on a large-scale design when evaluating outputs from RL agents.
\medskip

\bibliographystyle{abbrv}
\bibliography{bibliography}

\end{document}